\documentclass[aps,twocolumn]{revtex4}
\usepackage{amsmath}
\usepackage{graphicx}

\begin{document}

\title{High-order harmonic generation from diatomic molecules in
elliptically polarized driving fields: a generalized interference condition}
\author{T. Das$^{1}$, B. B. Augstein$^{1,2}$ and C. Figueira de Morisson Faria$^{1}$\\
$^{1}$Department of Physics and Astronomy, University College London,\\ Gower Street, London WC1E, 6BT, UK\\$^{2}$School of Chemistry, University of Leeds, Leeds LS2 9JT, UK}

\begin{abstract}
We investigate the influence of intense, elliptically polarized driving fields on high-order
harmonic spectra from aligned diatomic molecules. We derive a generalized two-center interference
condition for elliptically polarized fields, which accounts for $s-p$ mixing
and the orbital symmetry, within the strong-field and the single-active
electron approximation. We show that the non-vanishing ellipticity introduces
 an effective dynamic shift in the
angle for which the two-center interference maxima and minima occur, with
regard to the existing condition for linearly polarized fields. This shift
depends on the ratio between the field-dressed momentum components of the returning electron parallel and
perpendicular to the major ellipticity axis along each possible orbit.
Because of this dependence, we find that there will be a blurring in the
two-center interference minima, and that increasing ellipticity leads to
splitting in such patterns. These features are investigated in detail for $\mathrm{H}_2$ and $\mathrm{Ar}_2$.
\end{abstract}

\date{\today}
\maketitle
\affiliation{$^1$Department of Physics and Astronomy, University College London, Gower
Street, London WC1 6BT, UK}

\address{Department of Physics and Astronomy, University College London,\\
Gower Street, London WC1E 6BT, UK}

\section{Introduction}

Since the mid 1990s, elliptically polarized fields have been proposed as a
resource for controlling strong-field phenomena, such as high-order harmonic
generation (HHG), and its applications. Concrete examples are the production
of isolated attosecond pulses \cite%
{Corkum_1994,Ivanov_1995,Antoine_1996,Chang_2004}, and, more recently,
attosecond molecular imaging \cite{Kitzler_2007,Shafir_2009}. This control
is possible due to the fact that HHG owes its existence to a three-step
physical mechanism \cite{Corkum_1993}, in which an electron is freed via
multiphoton or tunnel ionization. Subsequently, it propagates in the
continuum and accumulates kinetic energy from the laser field on its return
to the parent ion. If it recombines with the ion it releases the energy in
the form of emitted high harmonic radiation. A typical high-order harmonic
spectrum exhibits a plateau, with harmonics of comparable intensities,
followed by a sudden cutoff, whose energy position roughly corresponds to
the maximal kinetic energy of a returning electron. Whether an electron will
return to the parent ion or not depends on the time it is born into the
continuum, and on its subsequent propagation. By an adequate choice of the
external-field parameters, such as its temporal profile and polarization,
one may steer the motion of the active electron in the continuum and control
how it returns to the core. As a direct consequence, one may manipulate
high-harmonic spectra.

This is the key idea behind polarization-gating techniques. An external
field with non-vanishing ellipticity will introduce a momentum component
perpendicular to the momentum an electron usually acquires from a linearly
polarized laser. This new degree of freedom may be controlled by modifying
the field ellipticity. For instance, lasers with changing ellipticity over
time were suggested in \cite{Corkum_1994,Ivanov_1995,Antoine_1996,Chang_2004}
as a way to produce isolated attosecond pulses. This was experimentally
realized in \cite{Sola_2006, Sansone_2006}, where the dependence of HHG on
the ellipticity of the driving pulses was used to create a temporal window
\cite{Sansone_2006} of linear polarization, for which, and only then, the
generation of extreme ultra violet (XUV) harmonics is possible. This
technique allows the generation of a broadband of XUV pulses with the
possibility of single cycle pulses. This is an improvement on the attopulses
produced through linearly polarized pulses, for which only the spectral
portion around the cut off can be used. Furthermore, polarization-gating
techniques allow a substantial increase in the intensity of the attosecond
pulses produced \cite{Kitzler_2007}.

Another important application of polarization-gated pulses is the attosecond
imaging of matter, in particular the reconstruction of molecular orbitals
\cite{Shafir_2009}. This imaging has been first realized with aligned
molecules in linearly polarized fields \cite{Itatani_2004}. However,
elliptically polarized fields exhibit a series of advantages. First, they
allow a greater degree of control of the angle with which an electron leaves
and returns to its parent ion \cite{Kitzler_2005,Shafir_2009}. Hence, in principle,
there is no necessity of aligning or rotating the molecule to be imaged.
Potentially, this provides access to degenerate orbitals, or molecules that
are difficult to align. Second, they allow molecular-orbital reconstruction
from a single-shot measurement. This may be useful for probing dynamic
processes in which space, energy and time coherence are important. Finally, by playing around with the field parameters, one may suppress or enhance the contributions of individual orbits along which the recolliding electron may return \cite{Brugnera_2011}.

In order, however, to be able to image molecules with elliptically polarized
fields, one must disentangle the imprints left by the field on the molecular
target from the features caused by the field itself. For linearly polarized driving fields, molecular imprints in HHG spectra have been widely studied, at least within the
single-active electron, single-active orbital approximation. For instance,
it is by now common knowledge that nodal planes cause a strong suppression
in HHG spectra if they are aligned parallel to the laser-field polarization
\cite{McFarland_2008}. Apart from that, the
high harmonic spectra from aligned molecules exhibit a
multi - slit like interference pattern, with pronounced maxima and minima,
which is dependent on the internuclear distance and the orientation of the molecule with respect to the
polarization of the laser field. This is a structural effect that results from the electron wave packet
recombining to spatially different centers. For the simplest scenario, i.e.,
a diatomic molecule, these interference patterns have been predicted since
the early 2000s \cite{Lein_3_2002} (for reviews see, e.g., \cite{Lein_2007}
and our recent publication \cite{Brad_2012}). Many of such studies have been
performed within the strong-field approximation (SFA), which has been
generalized to molecular systems (see, e.g., \cite{Kopold_1998,Madsen_2006,Madsen_2007,Chirila_2006,Faria_2007,Faria_2010,Brad_2011,Etches_2010,Odzak_2009,Odzak_2010}).
This approach has the particular advantage of allowing an almost entirely
analytic treatment and of providing a transparent physical interpretation in
terms of electron orbits, while retaining quantum-mechanical features such
as spatial and temporal interference. Recently, a generalized two-center
interference condition for high-order harmonic generation in
homonuclear diatomic molecules subjected to a linearly polarized laser field that accounts for the orbital geometry and
also $s-p$ mixing has been introduced \cite{Odzak_2009}.

Studies of the above-mentioned two-center interference for elliptically
polarized fields, however, are comparatively few. Most of the studies are focused on the
harmonic yield, as a function of the driving-field polarization \cite{Lein_JPB_2003}, or
on the ellipticity of the high-order harmonics as a way to probe the
anisotropy of a molecular medium \cite{Smirnova_2_2009,Etches_PRA_2010,Odzak_2011}. In particular, recent investigations have
shown that the minimum related to two-center interference becomes
increasingly blurred and appears to split if the ellipticity of the driving
field is increased \cite{Odzak_2010}. Therein, an interference condition for
the perpendicular molecular orientation was presented, which was different
along the major and the minor polarization axis of the driving field. The
focus of such papers, however, was on the vector character of the HHG
transition probabilities \cite{Odzak_2010}, and on the ellipticity of the
high-order harmonics \cite{Odzak_2011}. So far, the above-mentioned blurring
and splitting has not been addressed.

In this paper we present a two- center interference condition for diatomic
molecules in elliptically polarized fields. This condition is then tested in
HHG spectra computed employing the strong-field approximation for aligned diatomic molecules.
Throughout, we work within the single-active electron, single-active orbital approximation and assume the core to be frozen. One should note, however, that there may be also imprints caused by the dynamics of the core \cite{Smirnova_2009,Smirnova_2_2009}. Such effects will not be addressed in this work.

This article is organized as follows. In Sec.~\ref{Theory},
we generalize the molecular SFA to elliptically polarized driving fields. We
start from the standard SFA transition amplitude for high-order harmonic
generation, which is solved employing the steepest descent method (Sec.~\ref%
{SFAamplitude}). In Sec.~\ref{SFAelliptic}, this transition amplitude is explicitly written for elliptically polarized driving fields. Using such expressions, we derive
an analytic expression for two-center interference minima valid for
elliptical polarization, which contains an orbit-dependent, dynamic shift (Sec.~\ref{interference}). Subsequently, in Sec.~\ref{spectra}, we compute HHG spectra, and show that this dynamic shift is responsible for the blurring and splitting observed in \cite{Odzak_2010}. Finally,
in Sec. \ref{conclusions}, we provide our conclusions and summarize the main aspects of this work. We employ the length gauge and atomic units throughout.

\section{Model}
\label{Theory}
\subsection{Transition Amplitudes}
\label{SFAamplitude}
 The SFA transition amplitude for HHG \cite%
{Lewenstein_1994} is given by
\begin{align}  \label{Tamp}
M(\Omega)&=-i \int_{-\infty}^{\infty} dt \int_{-\infty}^{t} dt^{\prime 3}%
\mathbf{p} d^*_{rec}(\mathbf{p}+\mathbf{A}(t)) d_{ion}(\mathbf{p}+\mathbf{A}%
(t^{\prime}))  \notag \\
&\times e^{i S(t,t^{\prime },\Omega,\mathbf{p})} +c.c,
\end{align}
where
\begin{equation}
d_{ion}(\mathbf{p}) =\langle \mathbf{p}|H_I(t^{\prime })|\Psi_0 \rangle
\label{dion}
\end{equation}
and
\begin{equation}
d_{rec}(\mathbf{p})=\langle \mathbf{p}|\mathbf{r}| \Psi_0\rangle
\label{drec}
\end{equation}
are the ionization and recombination dipole matrix elements, respectively.
The semiclassical action
\begin{equation}  \label{action}
S(t,t^{\prime },\Omega,\mathbf{p})= -\frac{1}{2} \int^t_{t^{\prime }}[%
\mathbf{p}+\mathbf{A}(\tau)]^2 d\tau - I_p(t-t^{\prime}) + \Omega t
\end{equation}
describes the propagation of an electron of field-dressed momentum $\mathbf{p%
}+\mathbf{A}(\tau)$ in the continuum from the ionization time $t^{\prime }$
to the recombination time $t$. In the above-stated equations, $\mathbf{A}$
denotes the vector potential, $I_p$ the ionization potential, $\Omega$ the
harmonic frequency, and $H_I(t^{\prime})= \mathbf{r}\cdot \mathbf{E}%
(t^{\prime})$ the length-gauge interaction Hamiltonian at the ionization
time $t^{\prime}$.

In this work, we assume that all the influence of the molecular structure is
in the prefactors (\ref{dion}) and (\ref{drec}), and that only the highest
occupied molecular orbital (HOMO) contributes to the dynamics. These are the
most widely used assumptions within the molecular strong-field
approximation. Other assumptions, such as incorporating the structure of the
molecule in the action \cite{Kopold_1998,Chirila_2006,Faria_2007} or
employing models with more than one active orbital \cite%
{Patchkovskii_2006,Smirnova_2009,Smirnova_2_2009,Faria_2010}, have also been
used in the literature.

We calculate the transition amplitude (\ref{Tamp}) using the steepest
descent method. This implies that we solve Eq.~(\ref{Tamp}) by finding $%
t^{\prime}$, $t$ and $\mathbf{p}$ for which Eq.~(\ref{action}) is
stationary. This gives us the saddle-point equations
\begin{equation}  \label{t'saddle}
\frac{\partial S(t,t^{\prime },\mathbf{p})}{\partial t^{\prime }} =\frac{[%
\mathbf{p}+\mathbf{A}(t^{\prime})]^2}{2}+I_p=0,
\end{equation}
\begin{equation}  \label{psaddle}
\frac{\partial S(t,t^{\prime },\mathbf{p})}{\partial \mathbf{p}}
=\int^t_{t^{\prime }} d\tau[\mathbf{p}+\mathbf{A}(\tau)]=\mathbf{0}
\end{equation}
and
\begin{equation}  \label{tsaddle}
\frac{\partial S(t,t^{\prime },\mathbf{p})}{\partial t} =\frac{[\mathbf{p}+%
\mathbf{A}(t)]^2}{2}+I_p-\Omega=0.
\end{equation}
Physically, Eq.~(\ref{t'saddle}) expresses the conservation of energy for
the active electron upon tunnel ionization. Since tunneling has no classical
counterpart, this equation has no real solution. Eq.~(\ref{psaddle}) fixes
the intermediate momentum of the electron so that it returns to the site of
its release. In the present model, this is assumed to be the geometrical
center of the molecule, at $\mathbf{r}=0$. Finally, Eq.~(\ref{tsaddle})
gives the conservation of energy of the active electron upon recombination,
in which its kinetic energy is converted in a high-harmonic photon of
frequency $\Omega$. Throughout, when computing the transition probabilities associated with pairs of orbits, we employ a uniform approximation that treats each pair collectively. The transition probabilities associated with individual orbits are computed using the standard saddle-point approximation, which allows the orbits to be treated individually (for details see Ref.~\cite{Faria_2002}).

\subsection{Elliptically polarized fields}
\label{SFAelliptic}
 We will now assume that the external driving field is
elliptically polarized, i.e., made up of two linearly polarized orthogonal
laser fields. Throughout, we will adopt the subscripts ($||$) and ($\perp$)
to designate the momentum and field components parallel to the major and
minor polarization axis, respectively.

This implies that the time dependent electric field $\mathbf{E}(t)=-d\mathbf{%
A}(t)/dt$ and the vector potential $\mathbf{A}(t)$ may be written as
\begin{equation}
\mathbf{E}(t)=E_{\parallel}(t)\hat{\epsilon}_{\parallel}+E_{\perp}(t)\hat{%
\epsilon}_{\perp}
\end{equation}
and
\begin{equation}
\mathbf{A}(t)=A_{\parallel}(t)\hat{\epsilon}_{\parallel}+A_{\perp}(t)\hat{%
\epsilon}_{\perp},
\end{equation}
where the unit vector along the major and the minor polarization axis are
denoted by $\hat{\epsilon}_{\parallel}$ and $\hat{\epsilon}_{\perp}$,
respectively.

For this specific case, it is convenient to re-write the action as
\begin{align}
S(t,t^{\prime },\Omega ,\mathbf{p})& =-\frac{1}{2}\int_{t^{\prime
}}^{t}d\tau \lbrack p_{||}+A_{||}(\tau )]^{2}  \notag  \label{actionEP} \\
& -\frac{1}{2}\int_{t^{\prime }}^{t}d\tau \lbrack p_{\perp }+A_{\perp }(\tau
)]^{2}-I_{p}(t-t^{\prime })+\Omega t,
\end{align}%
and the saddle-point equations as
\begin{equation}
\frac{\partial S(t,t^{\prime },\mathbf{p})}{\partial t^{\prime }}=\frac{%
[p_{||}+A_{||}(t^{\prime })]^{2}}{2}+\frac{[p_{\perp }+A_{\perp }(t^{\prime
})]^{2}}{2}+I_{p}=0,  \label{tunnel}
\end{equation}%
\begin{equation}
\frac{\partial S(t,t^{\prime },\mathbf{p})}{\partial \mathbf{p}}%
=\int_{t^{\prime }}^{t}d\tau \lbrack \mathbf{p}_{||}+\mathbf{A}_{||}(\tau
)]+\int_{t^{\prime }}^{t}d\tau \lbrack \mathbf{p}_{\perp }+\mathbf{A}_{\perp
}(\tau )]=\mathbf{0},  \label{psaddleEP}
\end{equation}%
and
\begin{equation}
\frac{\partial S(t,t^{\prime },\mathbf{p})}{\partial t}=\frac{%
[p_{||}+A_{||}(t)]^{2}}{2}+\frac{[p_{\perp }+A_{\perp }(t)]^{2}}{2}%
+I_{p}-\Omega =0,  \label{rec}
\end{equation}%
respectively. From Eq.~(\ref{psaddleEP}) we obtain an equation for the
stationary momentum for elliptically polarized fields,
\begin{equation}
\mathbf{p}_{\mathrm{st}}=p_{\mathrm{st}\parallel }\hat{\epsilon}_{\parallel
}+p_{\mathrm{st}\perp }\hat{\epsilon}_{\perp },  \label{pstat}
\end{equation}%
where
\begin{equation}
p_{st\parallel }=\frac{-1}{t-t^{\prime }}\int_{t^{\prime }}^{t}A_{\parallel
}(\tau )d\tau   \label{pstatpar}
\end{equation}%
and
\begin{equation}
p_{st\perp }=\frac{-1}{t-t^{\prime }}\int_{t^{\prime }}^{t}A_{\perp }(\tau
)d\tau .  \label{pstatperp}
\end{equation}%
Eq.~(\ref{pstat}) implies that, within the saddle-point approximation, the
dynamics will be concentrated along the ellipticity plane.

Note also that, for elliptically polarized fields, differentiation of Eq.~(\ref{t'saddle}) with regard to $t^{\prime}$  and of Eq.~(\ref{tsaddle}) with regard to $t$ gives
\begin{equation}
[\mathbf{p}+\mathbf{A}(t^{\prime})]\cdot \mathbf{E}(t^{\prime})=0
\label{perpclass1}
\end{equation}
and
\begin{equation}
[\mathbf{p}+\mathbf{A}(t)] \cdot \mathbf{E}(t)=0,
\label{perpclass2}
\end{equation}
respectively. Eqs.~(\ref{perpclass1}) and (\ref{perpclass2}) hold in the classical limit $I_p\rightarrow 0$.  Physically, these equations state that the electron velocity $\mathbf{p}+\mathbf{A}(\tau)$ with $\tau=t,t^{\prime}$ must be perpendicular to the electric field at the instant of ionization and recombination. This condition has been employed in Ref.~\cite{Goreslavskii2004}, in the context of above-threshold ionization.

\subsection{Interference Condition}
\label{interference}
 In order to derive the interference condition for
elliptically polarized fields, we will focus on the explicit expression for
the recombination prefactor $d_{rec}$. The ionization prefactor $d_{ion}$
will influence the overall intensity in the spectrum, and is not relevant
for a qualitative discussion of two-center interference effects \cite%
{Faria_2007}.

Within our model, we represent the HOMO by a linear combination of atomic
orbitals (LCAO) and neglect the motion of the nuclei. Hence, the HOMO
wavefunction $\Psi _{0}(\mathbf{r})$ reads
\begin{equation}
\Psi _{0}(\mathbf{r})=\hspace*{-0.2cm}\sum_{a}c_{a }\hspace*{-0.1cm}\left[ \psi _{a }\hspace*{-0.1cm}\left( \mathbf{r}%
+\frac{\mathbf{R}}{2}\right) \hspace*{-0.1cm}+\hspace*{-0.1cm}(-1)^{\ell _{a }-m_{a }+\lambda
_{a }}\psi _{a }\hspace*{-0.1cm}\left( \mathbf{r}-\frac{\mathbf{R}}{2}\right) %
\right] \hspace*{-0.1cm},  \label{HOMOwf}
\end{equation}%
where $\psi _{a }(\mathbf{r})$ are the atomic orbitals, \textbf{R} is
the internuclear distance, $\ell_{a} $ is the orbital quantum number and $m_{a}$ is the
magnetic quantum number. The indices $\lambda_{a} =m_{a}$ correspond to gerade (g) and $%
\lambda_{a} =m_{a}+1$ to ungerade (u) orbital symmetry, respectively.

Below we extend the two center interference condition for linearly polarized
light given in Ref.~\cite{Odzak_2009} to elliptically polarized light. We first
consider the dipole matrix element $d_{rec}(\mathbf{p}+\mathbf{A}(t))$ for
the wavefunction (\ref{HOMOwf}). This gives
\begin{eqnarray}
d_{rec}(\mathbf{p}(t))\hspace*{-0.1cm}&=\hspace*{-0.1cm}&\sum_a%
\hspace*{-0.15cm}c_a\left[ e^{i\mathbf{p}(t)\cdot \frac{\mathbf{R}}{2}%
}+(-1)^{\ell_a-m_a+\lambda_a}e^{-i\mathbf{p}(t)\cdot\frac{\mathbf{R}}{2}} \right]  \notag \\
&& \times i\partial_{\mathbf{p}(t)}\psi_{a}(\mathbf{p}(t)),
\end{eqnarray}
where $\mathbf{p}(t)=\mathbf{p} + \mathbf{A}(t)$ and
\begin{equation}
\psi_{a}(\mathbf{p}(t))=\frac{1}{(2\pi)^{3/2}}\int d^3r \psi_{a}(%
\mathbf{r})\exp[-i \mathbf{r} \cdot \mathbf{p}(t)].
\end{equation}
In the above-stated equations, the terms related to the lack of
orthogonality between the bound states and the continuum introduced by the
SFA have been neglected (see Refs.~\cite{Smirnova_2007,Faria_2007} for details). The quantity of
interest is $d^*_{rec}(\mathbf{p}(t)\cdot \mathbf{E}(t))$ along the
field-polarization direction. Explicitly,
\begin{eqnarray}
d^*_{rec}(\mathbf{p}(t)\cdot \mathbf{E}(t))\hspace*{-0.1cm}&=\hspace*{-0.1cm}%
&\sum_a\hspace*{-0.1cm}c_a\hspace*{-0.1cm}\left[ e^{-i\mathbf{%
p}(t)\cdot \frac{\mathbf{R}}{2}}+(-1)^{\ell_a-m_a+\lambda_a}e^{i\mathbf{p}(t)\cdot \frac{\mathbf{R}}{2}}\right]
\notag \\
&& \times (-i) \sum_b \partial_{p_{b}(t)}\psi^*_a(\mathbf{p}%
(t))E_{b}(t),  \label{drec2}
\end{eqnarray}
where b= $||$, $\perp $ indicate the components along the major and minor
polarization axis.

Following the procedure in Ref.~\cite{Odzak_2009} for linearly polarized laser
fields, Eq.~(\ref{drec2}) can be rewritten as
\begin{eqnarray}
d_{rec}^{\ast }(\mathbf{p}(t)\cdot \mathbf{E}(t)) &=&\cos \left( \mathbf{p}%
(t)\cdot \frac{\mathbf{R}}{2}\right) A_{+}  \notag \\
&&+iA_{-}\sin \left( \mathbf{p}(t)\cdot \frac{\mathbf{R}}{2}\right) ,
\label{interf}
\end{eqnarray}%
where
\begin{equation}
A_{\pm }=\sum_a c_a\left[ (-1)^{\ell _a+m_a+\lambda _a}\pm 1\right] \eta (\mathbf{p}+\mathbf{A}(t),t)
\end{equation}%
and
\begin{equation}
\eta (\mathbf{p},t)=-i\left[ \partial _{p_{\parallel }}\psi _a^{\ast }(\mathbf{p})E_{\parallel }(t)+\partial _{p_{\perp }}\psi _a^{\ast }(\mathbf{p})E_{\perp }(t)\right] .
\end{equation}%
Note that, because there is an
electric field component $E_{\perp }(t)$ and a field-dressed momentum
component $p_{\perp }(t)$ along the minor polarization axis, the function $%
\eta (\mathbf{p},t)$, the s-p mixing embedded in $A_{\pm }$, will
be different from the expressions obtained in Ref.~\cite{Odzak_2009} for linear polarization. Re-writing Eq.~(\ref{interf}) as
\begin{equation}
d_{rec}^{\ast }(\mathbf{p}(t)\cdot \mathbf{E}(t))=\sqrt{A_{+}^{2}-A_{-}^{2}}%
\sin \left[ \mathbf{p}(t)\cdot \frac{\mathbf{R}}{2}+\alpha \right] ,
\label{interfb}
\end{equation}%
where $\alpha =\arctan {\frac{-iA_{+}}{A_{-}}}$, we expect interference
minima at
\begin{equation}
\alpha +\mathbf{p}(t)\cdot \frac{\mathbf{R}}{2}=n\pi .
\end{equation}%
For elliptically polarized fields we have
\begin{equation}
\mathbf{p}(t)\cdot \frac{\mathbf{R}}{2}=p_{||}(t)\frac{R}{2}\cos {\theta _{L}%
}+p_{\perp }(t)\frac{R}{2}\sin {\theta _{L}}
\end{equation}%
where $\theta _{L}$ is the angle between the molecular internuclear and the
major polarization axis. Using
\begin{equation}
\sqrt{(\mathbf{p}+\mathbf{A}(t))^{2}}\cos {\beta }=[p_{||}+A_{||}(t)]\cos {%
\theta _{L}}+[p_{\perp }+A_{\perp }(t)]\sin {\theta _{L}},
\end{equation}%
where $\sqrt{(\mathbf{p}+\mathbf{A}(t))^{2}}=\sqrt{(p_{\parallel
}+A_{\parallel }(t))^{2}+(p_{\perp }+A_{\perp }(t))^{2}}$ and calling
\begin{equation}
\frac{\lbrack p_{||}+A_{||}(t)]}{\sqrt{(\mathbf{p}+\mathbf{A}(t))^{2}}}=\cos
\zeta (t,t^{\prime})   \label{xi1}
\end{equation}%
\begin{equation}
\frac{\lbrack p_{\perp }+A_{\perp }(t)]}{\sqrt{(\mathbf{p}+\mathbf{A}(t))^{2}%
}}=\sin \zeta (t,t^{\prime}),  \label{xi2}
\end{equation}%
we obtain
\begin{equation}
\sqrt{(\mathbf{p}+\mathbf{A}(t))^{2}}\frac{R}{2}\cos ({\theta _{L}-\zeta(t,t^{\prime}) })%
=n\pi -\alpha ,  \label{min}
\end{equation}%
where
\begin{equation}
\zeta (t,t^{\prime })=\arctan \left[ \frac{p_{\perp }+A_{\perp }(t)%
}{p_{||}+A_{||}(t)}\right] .\label{shift}
\end{equation}

Physically, this equation demonstrates that a field of non-vanishing
ellipticity introduces an effective shift $\zeta (t,t^{\prime })$ in the
alignment angle $\theta _{L}$ at which the interference minimum in the
harmonic spectrum occurs, with regard to the linearly polarized case. Using
Eq.~(\ref{min}) and Eq.~(\ref{tsaddle}) we find that the destructive
interference leading to minima in the harmonic spectrum is determined by the
expression
\begin{equation}
\Omega =\frac{2[n\pi -\alpha ]^{2}}{R^{2}\cos ^{2}(\theta _{L}-\zeta
(t,t^{\prime }))}+I_{p}  \label{intcon}
\end{equation}

From Eqs.~(\ref{xi1}) and (\ref{xi2}) it is clear that the value of $\zeta$
depends upon the field dressed momentum components $p_{\parallel
}+A_{\parallel }(t)$ and $p_{\perp }+A_{\perp }(t)$ of the returning
electron, and hence on its return time $t$ along each orbit. Furthermore, $%
p_{\parallel }$ and $p_{\perp }$ are functions of the return and ionization
times $t$ and $t^{\prime }$ according to the saddle-point Eqs.~(\ref%
{pstatpar}) and (\ref{pstatperp}). Therefore, the location of the minimum in
the harmonic spectrum given by Eq.~(\ref{intcon}) is dependent on the electron orbit, i.e., the elliptical polarization introduces a dynamical shift.
This implies that, whereas in the case of linearly polarized fields there is a clear harmonic at which destructive
interference occurs for any given alignment angle, and the interference condition is purely structural,
 in the case of an elliptically polarized field, we expect to find
minima in various places in the harmonic spectrum depending upon the
intermediate momentum components. As the overall spectrum is constructed
from the coherent sum of a large number of electron orbits, the above
condition is likely to result in blurring and in splitting in the two-center
minima found in the harmonic spectrum. In our computations, we have included
up to the three shortest pairs of orbits. The longer pairs have little effect
in the harmonic spectrum due to wave-packet spreading \cite{Faria_2010}.

\section{High-harmonic spectra}
\label{spectra}
In the results that follow, we analyze the interference condition derived in
the previous section, for an elliptically polarized field of the form
\begin{equation}
\mathbf{E}(t)=\frac{ E_{0}}{\sqrt{1+\xi ^{2}}}\left[\sin (\omega t)\hat{\epsilon}%
_{\parallel }+\xi \sin (n\omega t-2\pi \phi )\hat{\epsilon}%
_{\perp }\right] ,  \label{field}
\end{equation}%
where the frequency ratio is chosen as $n=2$. This corresponds to a two-color field composed of a monochromatic wave of frequency $\omega $ along the major polarization axis and of its second
harmonic along the minor axis, respectively. Two-color fields with elliptical polarization have been recently employed in \cite{Kitzler_2005,Kitzler_2007,Shafir_2009,Brugnera_2011}.

In Eq.~(\ref{field}), $\xi $ determines the relative strength of
the field component along the minor polarization axis with regard to its
component along the major axis, and $\phi $ is a phase factor which
determines the time delay between both waves. The field has been normalized
in such a way that the overall time-averaged intensity $\langle \mathbf{E}^2(t) \rangle_t$ remains constant. For a monochromatic field ($n=1$), this implies that the total ponderomotive energy $U_p=\langle A_{\parallel}^2(t)\rangle_t/2+\langle A_{\perp}^2(t)\rangle_t/2$ remains constant as well. For two-color fields, this condition implies that $U_p$ will decrease with the driving-field ellipticity \footnote{In the bichromatic case, constant $U_p$ requires division of the field amplitude by an overall factor $\sqrt{1+\xi ^{2}/4}$. This factor, however, rules out an overall constant time-averaged intensity.}.

The highest occupied molecular orbital (HOMO) is constructed using
Gaussian-type orbitals computed with GAMESS-UK \cite{GAMESS}. Only $s$ and $p
$ orbitals are included in the basis sets employed in this work.

%In contrast, the HOMO of  $N_{2}$ is a 3$\sigma _{g}$
%orbital with a high degree of $s-p$ mixing. Hence, it will be useful to
%investigate how the generalized interference condition may be applied to
%such scenarios. \ Throughout, we will consider the harmonic spectra in the
%direction of the major axis of the polarization ellipse.
%
\subsection{Testing the interference condition}

As a starting point, we will focus on whether the effective shift $\zeta
(t,t^{\prime })$ can be identified and whether it agrees with Eq.~(\ref{intcon}). For that purpose, we will compute the transition probabilities $|M(\omega)|^2$ associated with individual orbits along which the active electron returns to the core, starting from the dominant, shortest pair of orbits.
These orbits are well known in the literature as \textquotedblleft the long
orbit" and \textquotedblleft the short orbit" \cite{Antoine_1996_PRL}, and correspond to electron excursion times of the order of three quarters of a field cycle. Throughout, we will classify all electron orbits according to increasing ionization times employing positive integers. For simplicity, we will first consider $\mathrm{H}_{2}$ as a target. Since its HOMO is a $1\sigma_g$ orbital composed of $s$ orbitals only, $\mathrm{H}_{2}$ is very useful for investigating whether Eq.~(\ref{intcon}) holds. The overall field intensity has been taken to be the same as in Ref.~\cite{Lein_2002}.

\begin{figure}[tbp]
\noindent\includegraphics[scale=0.3]{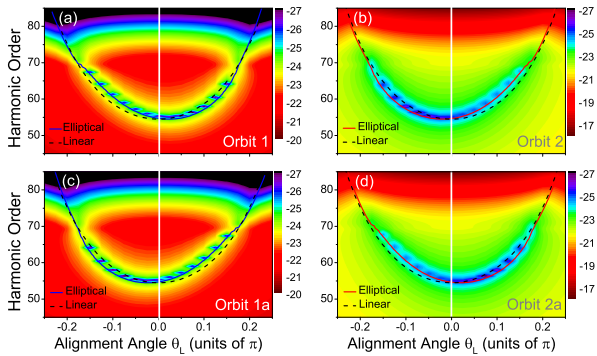}
\caption{(Color online) Harmonic spectra along the major polarization axis as functions of
the alignment angle $\protect\theta _{L}$ \ for $H_{2}$ ($I_p=$0.5 a.u. and internuclear separation $R=1.4$ a.u.) in an elliptical
field described in Eq.~(\protect\ref{field}) with $n=2$, $\protect\omega =0.057$
a.u., $I$=5$\times 10^{14}\mathrm{Wcm}^{-2}$, $\protect\xi =0.3$ and time delay $%
\protect\phi =0.2$. Panels (a) and (c) show the spectra for the long
electron orbits 1 and 1a starting in the first and second half cycle,
respectively, while  panels (b) and (d) exhibit the spectra obtained for the
short orbits 2 and 2a starting in the first and second half cycle,
respectively.  The generalized interference condition (\protect\ref{intcon})
is indicated by the solid lines in the figure, whereby we have just considered the real parts $\mathrm{Re}[\zeta(t,t^{\prime})]$ of the time-dependent shifts. For comparison, we plot the
two-center interference condition for linearly polarized fields as the
dashed lines. The central white lines indicate vanishing alignment angle $\theta_L=0$. The harmonic yield is given in a logarithmic scale. The increase in the harmonic yields after the cutoff observed in the right panels are related to a breakdown of the standard saddle-point approximation for the short orbits (for details see Ref.~\protect\cite{Faria_2002}).}
\label{Fig1}
\end{figure}

In Fig.~\ref{Fig1}, we display these contributions as functions
of the alignment angle $\theta _{L}$ between the internuclear axis and the
major polarization axis. Orbits 1 and 2 [panels (a) and (b), respectively] start in the first half cycle of
the driving field, slightly after the first field peak, and return close to the field crossing at the end of the first field cycle $t=T=2\pi/\omega$.
In the lower panels of the figure, we display the contributions from orbits 1a and 2a [panels (c) and (d), respectively], whose start and return times are
 displaced by half a cycle with regard to those of orbits 1 and 2.
Throughout, the two-center interference conditions are indicated, both for linearly and elliptically polarized fields (dashed and solid lines, respectively). In the elliptic case, we have considered the real parts of the dynamic shift, i.e.,  $\mathrm{Re}[\zeta(t,t^{\prime})]$ when plotting the two-center minimum. We have verified that this approximation is accurate enough for individual orbits, as $\mathrm{Im}[\zeta(t,t^{\prime})]$ is vanishingly small in the harmonic ranges of interest.

As an overall feature, we observe an excellent agreement between Eq.~(\ref{intcon}) and the outcome of the SFA computations, with the two-center minimum varying from orbit to orbit. Moreover, in contrast to what happens for linearly polarized fields, the minimum is no longer symmetric upon $\theta _{L}\rightarrow -\theta _{L}$. These features can be explained in terms of the time dependence of the effective shift $\zeta (t,t^{\prime })$. For a specific
orbit, the times $t,t^{\prime}$ will only vary with the harmonic energy $\Omega $. Hence,
shifting $\theta_L$ to $-\theta_L$ does not imply shifting $\zeta(t,t^{\prime})$ to $-\zeta(t,t^{\prime})$,
and the above-mentioned symmetry will be broken. Furthermore, because $t$ and $t^{\prime}$ are orbit dependent, we observe different shifts $\zeta(t,t^{\prime})$ for different orbits. In fact, for orbits 1 and 2a, the shifts displace the interference minimum to the right, while for orbits 1a and 2, this displacement is to the left.

Interestingly, the shifts observed for orbits $1$ and $2$ are the mirror image of those obtained for orbits 1a and 2a, respectively.  This is due to the specific behavior of the two-color driving field for $t\rightarrow t \pm T/2$, where $T/2=\pi/\omega$. In this case, $A_{\parallel}(t\pm T/2)=-A_{\parallel}(t)$, and $A_{\perp}(t\pm T/2)=A_{\perp}(t)$. Hence, direct inspection of Eq.~(\ref{shift}) shows that $\zeta (t,t^{\prime })=-\zeta (t\pm T/2,t^{\prime }\pm T/2)$. For a monochromatic elliptically polarized field, i.e., $n=1$ in Eq.~(\ref{field}), in contrast, $\zeta(t,t^{\prime})=\zeta (t\pm T/2,t^{\prime }\pm T/2)$, i.e., the shift will remain invariant if the ionization and return times are displaced in half a cycle.

\begin{figure}[tbp]
\begin{center}
\includegraphics[scale=0.25]{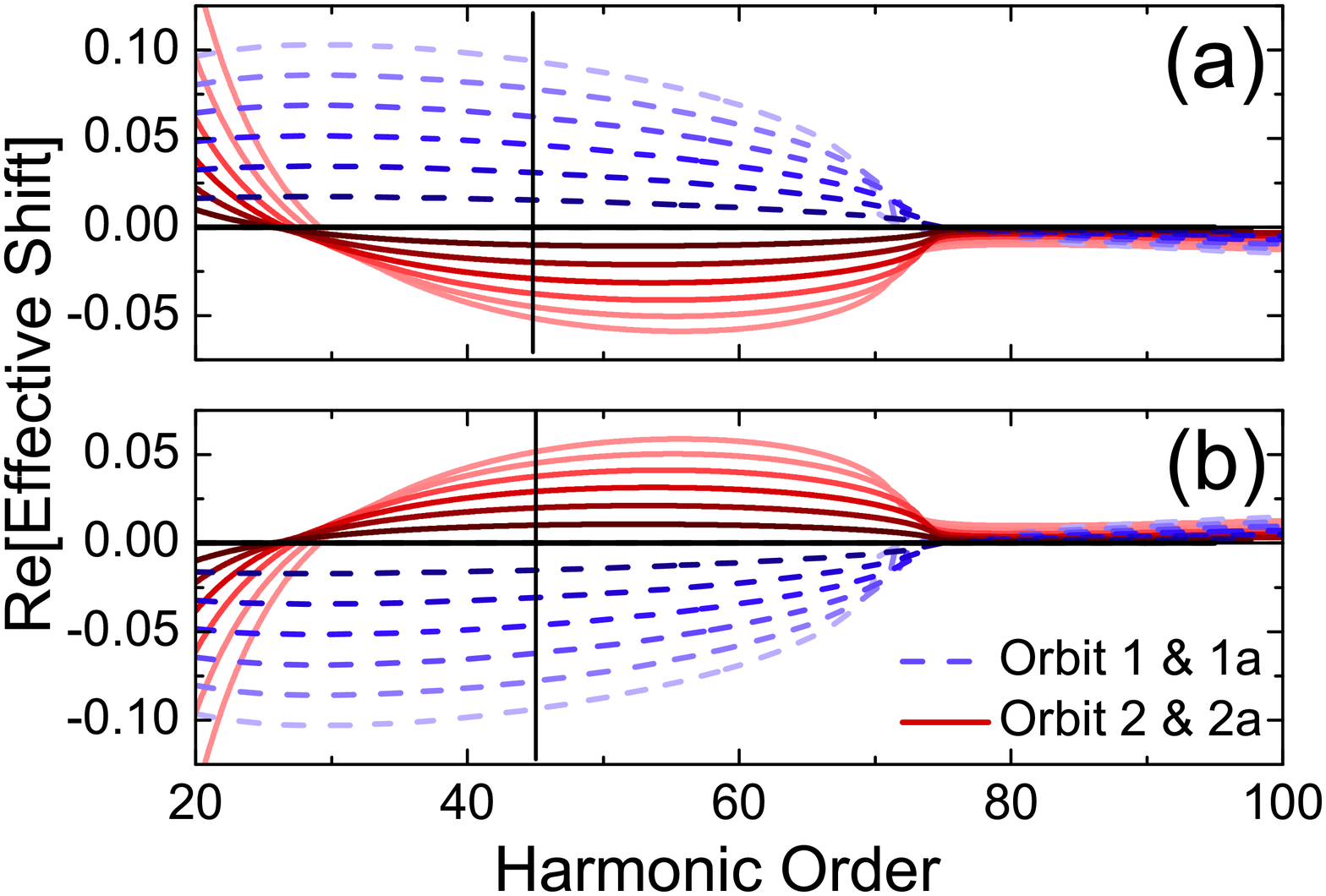}
\end{center}
\caption{ (Color online) Real parts of the effective shifts $\protect\zeta (t,t^{\prime })$ as functions of the harmonic order computed for orbits 1 and 2 [panel (a)] and orbits 1a and 2a [panel (b)], using $\mathrm{H}_2$ in two-color laser fields of increasing ellipticity and the same relative phase $\phi$, intensity and frequency as in Fig.~\ref{Fig1}. The ellipticities have been increased from $\xi=0$ to $\xi=0.3$ in increments of $\Delta \xi=0.05$. A lighter color indicates a higher ellipticity. For clarity, the harmonic range in which Fig.~\ref{Fig1} starts is indicated by a black vertical line and a vanishing shift is indicated by a horizontal black line. The dashed lines refer to the orbits $1$ and 1a, while the solid lines correspond to orbits $2$ and 2a. }
\label{Fig2}
\end{figure}

The above-stated observation is confirmed by Fig.~\ref{Fig2}, in
which the real parts of the effective shifts $\zeta (t,t^{\prime })$ are
plotted for driving fields of increasing ellipticity. The case considered in the previous figure, i.e., $\xi=0.3$, is given by the outer curves. For the harmonic range considered in Fig.~\ref{Fig1}, i.e., $45\leq\Omega/\omega\leq90$, $\mathrm{Re}[\zeta (t,t^{\prime })]>0$  for orbits $1$ and 2a. This is consistent with the fact that the interference minimum shifts to the right for both orbits [see Figs.~\ref{Fig1}(a) and (d)]. Indeed, when subtracted from a positive alignment angle $\theta_L$, a positive shift will displace the interference condition (\ref{intcon}) towards lower harmonics. For $\theta_L<0$, on the other hand, subtracting a positive shift will bring the minimum towards higher energies. Beyond the cutoff, the real parts of the shifts decrease substantially. Consequently, the interference condition will approach that obtained for a linearly polarized field. This is clearly seen in Fig.~\ref{Fig1}, for harmonic order $\Omega/\omega\ge 69$. A similar analysis can be performed for orbits $2$ and 1a, for which the u-shaped minimum is displaced to the left in Fig.~\ref{Fig1}(b) and (c). In this latter case, $\mathrm{Re}[\zeta(t,t^{\prime})]<0$ in the harmonic range of interest. Note, however, that there is a small residual shift beyond the cutoff, whose real part is negative for the orbits starting at the first half cycle, and positive for those starting at the second half cycle [Figs.~\ref{Fig2}(a) and (b), respectively]. Hence, the minimum for elliptic polarization will approach its counterpart for linearly polarized fields from the right in Figs.~\ref{Fig1}(a) and (b), and from the left in Figs.~\ref{Fig1}(b) and (d).
% This should be the new Fig. 3
\begin{figure}[tbp]
\noindent\includegraphics[scale=0.3]{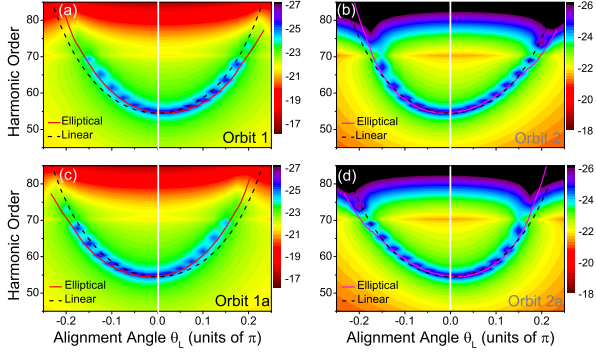}
\caption{(Color online) Transition probabilities associated with individual orbits for $\mathrm{H}_2$ in an elliptically polarized field with the same parameters as in in Fig.~\ref{Fig1}, but time delay $\phi=0$ between the $\omega$ and the $2\omega$ waves. Panels (a) and (c) correspond to the long orbits $1$ and 1a, while panels (b) and (d) give the contributions of the short orbits $2$ and 2a. The interference minima for linear and elliptically polarized fields are indicated by the solid and dashed lines in the figure.  The increase in the harmonic signal after the cutoff observed in the left panels is related to a breakdown of the standard saddle-point approximation for the long orbits (for details see Ref.~\protect\cite{Faria_2002}). The harmonic yields are displayed in a logarithmic scale.}
\label{Fig3}
\end{figure}
%%%%%%%%%%%%%%%%%%%%%%%%%%%%%%%%%%%%%
\begin{figure}[tbp]
\noindent\includegraphics[scale=0.25]{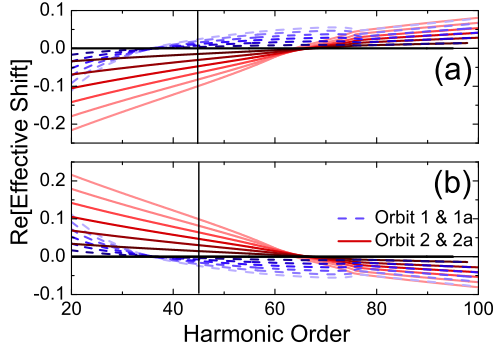}
\caption{(Color online) Real parts of the shifts $\zeta(t,t^{\prime})$ computed for a two-color elliptically polarized field (\ref{field}) with $n=2$ and $\phi=0$. Panels (a) and (b) refer to the orbits released in the first and second half cycle, respectively. The remaining molecular and field parameters are the same as in Fig.~\ref{Fig2}. }
\label{Fig4}
\end{figure}

One should note, however, that these shifts are strongly dependent on the time delay between the low- frequency and high-frequency waves. An example is provided in Fig.~\ref{Fig3}, for which both driving waves are in phase, i.e., $\phi=0$.  The minima for the dominant orbits $1$, 1a, $2$ and 2a once more follow the generalized interference condition (\ref{intcon}). The curves, however, are markedly different from those displayed in Fig.~\ref{Fig1}. A noteworthy feature is that there are now large residual shifts beyond the cutoff. This is explicitly shown in Fig.~\ref{Fig3}. There is once more a very good agreement between Eq.~(\ref{intcon}) and the minima encountered, but there are large residual shifts beyond the cutoff.

This agrees with Fig.~\ref{Fig4}, in which the real parts of the shifts are displayed for the dominant orbits and $\phi=0$. In contrast to what has been observed in Fig.~\ref{Fig2}, $\mathrm{Re}[\zeta (t,t^{\prime})]$ has a non-vanishing value at the cutoff. For instance, for orbits 1 and 2, the residual shift at the cutoff is positive. However, if the electron returns half a cycle later, i.e., along orbit 1a or 2a, this shift is negative. This is expected as the major component $A_{\parallel}(t)$ and $A_{\parallel}(t \pm T/2)$ have different signs.
\begin{figure}[tbp]
\noindent\includegraphics[scale=0.3]{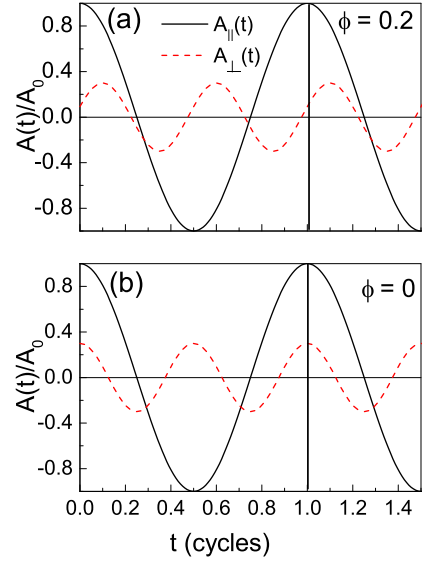}
\caption{(Color online) Schematic representation of the major and minor components of the vector potential $A(t)$ for ellipticity $\xi=0.3$, frequency ratios $1:2$ [$n=2$ in Eq.~(\ref{field})], and relative phases $\phi=0.2$ and $\phi=0$ [panels (a) and (b), respectively]. The electron return time at $t=2\pi/\omega$ is indicated by the thick black lines in the figure. For simplicity, all fields have been normalized  to the vector potential amplitude $A_0=E_0/\omega$.}
\label{Fig5}
\end{figure}
%%%%%%%%%%%%%%%%%%%%%%%%%%%%%%%%%%%%%%%%%%%%%%%%%%%%%%%%%%%%%%%%%%%%%%%%%%%%%%%%%%%%%%%%%%%%%%%%%%
%%%%%%%%%%%%%%%%%%%%%%%%%%%%%%%%%%%%%%%%%%%%%%%%%%%%%%%%%%%%%%%%%%%%%%%%%%%%%%%%%%%%%%%%%%%%%%
\begin{figure*}[tbp]
\noindent\includegraphics[scale=0.65]{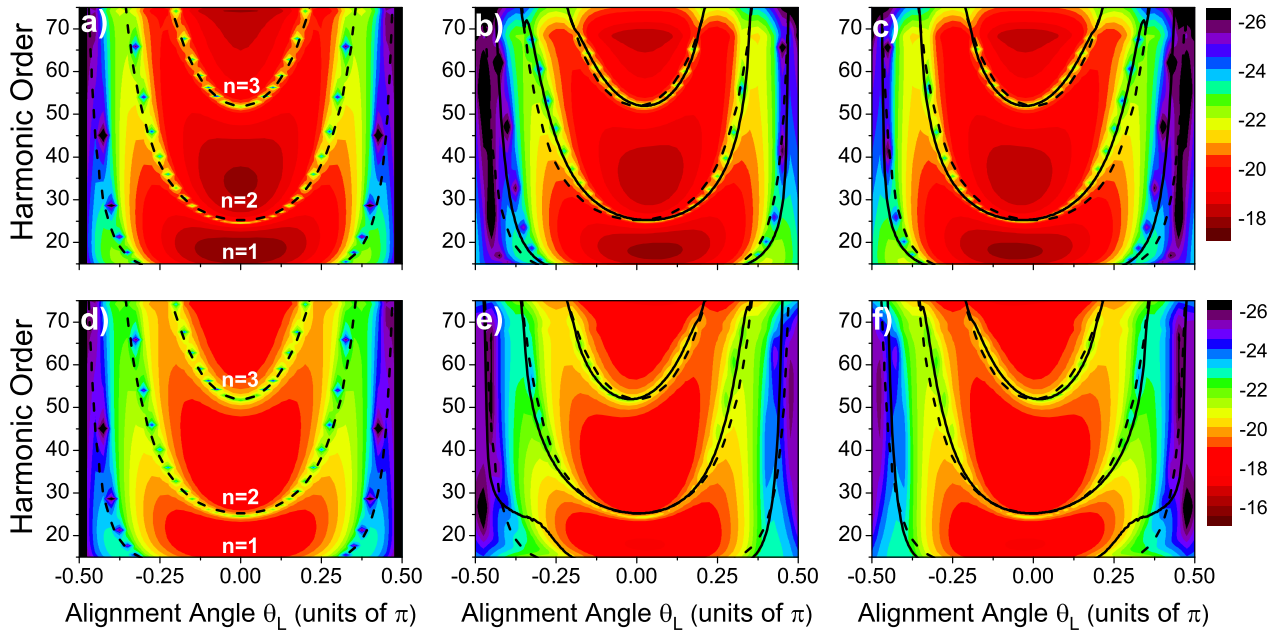}
\caption{ (Color online) Harmonic spectra along the major polarization axis computed for individual orbits as functions of
the alignment angle $\protect\theta _{L}$ for $\mathrm{Ar}_{2}$ (ionization potential $I_p=0.58$ a.u. and internuclear separation $R=7.2$ a.u.). For comparison, the individual-orbit contributions obtained for linear polarization are displayed in the far left panels (a) and (d), while in the middle and far right panels (b), (c), (e) and (f) the same $\omega-2\omega$ elliptically polarized field as in Fig.~\ref{Fig1} has been employed. Panels (b) and (c) exhibit the contributions from the long orbits 1 and 1a, while panels (e) and (f) depict the contributions from the short orbits 2 and 2a. For the upper and lower panels, the orbits start at the first and second half cycle, respectively.  The interference conditions for elliptically and linearly polarized fields are indicated as the solid and dashed lines in the figure, respectively. The harmonic yield is given in a logarithmic scale. The increase in the harmonic yields after the cutoff observed in panels (e) and (f) are related to a breakdown of the standard saddle-point approximation for the short orbits (for details see Ref.~\protect\cite{Faria_2002}).}
\label{Fig6}
\end{figure*}

 The behavior with the time delay $\phi$ may be understood if one takes into consideration that this phase difference has a strong influence on the velocity $p_{\perp}+A_{\perp}(t)$ of the electron upon return along the minor polarization axis. We have verified that, for a wide range of phases $\phi$, including $\phi=0$ and $\phi=0.2$, the electron return times are practically identical to those obtained for linearly polarized fields. Thus, at the cutoff, the electron will return near a crossing of the electric field $E_{\parallel}(t)$ along the major polarization axis. If $\phi=0.2$, the amplitude $|E_{\perp}(t)|$ will be close to its maximum. This implies that $|A_{\perp}(t)|/A_0\ll 1 $. Hence, $\mathrm{Re}[\zeta(t,t^{\prime})]$ is very small for the harmonics at and beyond the cutoff. On the other hand, if $\phi=0$, the perpendicular component  $A_{\perp}(t)/A_0=\pm 1$ for the cutoff return times. This implies that the residual shifts will be large.

 This can be seen in Fig.~\ref{Fig5}, where we provide an illustration of the vector potentials $A_{\parallel}(t_c)$ and $A_{\perp}(t_c)$ for the return times at a crossing. For $\phi=0.2$ [Fig.~\ref{Fig5}(a)], the vector potential  $A_{\perp}(t_c)$ is very small, and so is the shift at and beyond the cutoff. There is, however, a residual shift as the vector potential is not exactly zero.
For $\phi=0$, the transverse vector potential  $A_{\perp}(t_c)=\xi E_0/(2\omega)$ is at is maximum at $t=t_c$, as shown in Fig.~\ref{Fig5}(b), so that the transverse velocity of the electron upon return will be non-vanishing. Hence, at and beyond the cutoff $\mathrm{Re}[\zeta(t,t^{\prime})]\neq 0$. This will leave large residual shifts beyond the cutoff, as shown in the previous figures.

In order to see the behavior outlined in Fig.~\ref{Fig2} more clearly, it is desirable to seek a parameter range for which several minima are present over a wide harmonic energy range. This can be achieved by choosing a target with a large equilibrium internuclear distance, such as $\mathrm{Ar}_2$. The spectra computed for this target using individual orbits is displayed in Fig.~\ref{Fig6}, for the same driving field as in Figs.~\ref{Fig1} and \ref{Fig2}. For each panel, one may identify three interference minima. The lowest-order minimum spans the whole harmonic range displayed in Fig.~\ref{Fig2}, the intermediate minimum starts at approximately $\Omega=30\omega$, and the highest minimum covers similar harmonic frequencies to those studied in Fig.~\ref{Fig1}. The figure shows very distinct behaviors for the long and short orbits. For the long orbits there is a monotonic shift, either to the right [Fig.~\ref{Fig6}(b)], or to the left [Fig.~\ref{Fig6}(c)], while for the short orbits the sign of $\mathrm{Re}[\zeta(t,t^{\prime})]$ varies. As a direct consequence, the elliptical minima``wiggle" around their linear counterparts.  For example, for orbit 2 [Fig.~\ref{Fig6}(e)], there is a shift to the right for harmonic frequencies $\Omega\lesssim 30\omega$ in the two lower minima. Around this harmonic energy, the minimum crosses that obtained for linear polarization, and moves to the left. This is consistent with the behavior of the red solid curves in Fig.~\ref{Fig2}(a). For orbit 2a, the minimum follows the red curves in Fig.~\ref{Fig2}(b), i.e., they are the mirror image of those in Fig.~\ref{Fig6}(e) with regard to the shift $\theta_L \rightarrow -\theta_L$. This is explicitly shown in Fig.~\ref{Fig6}(f). Once more, beyond the cutoff the elliptical and the linear minima approach each other for $\phi=0.2$. In general, the outcome of the strong-field approximation follows the minima predicted by Eq.~(\ref{intcon}) reasonably well. An exception is, however, the interference minimum $n=1$ obtained for the short orbits in very low ($\Omega<20\omega$) and very high (i.e., beyond the cutoff) harmonic ranges [see Figs. \ref{Fig6}(e) and (f)]. These discrepancies are possibly due to the fact that, in these regions,the imaginary parts $\mathrm{Im}[\zeta(t,t^{\prime})]$ increase considerably for orbits $2$ and 2a. Thus, the approximation employed in the figure ceases to be accurate. Nevertheless, we have verified that the analytic condition (\ref{intcon}) is also valid in this energy region for $\xi\leq0.2$ (not shown).

\subsection{Coherent superpositions of orbits}
\label{Coherent}
\begin{figure}[tbp]
\noindent\hspace*{-0.5cm}\includegraphics[scale=0.3]{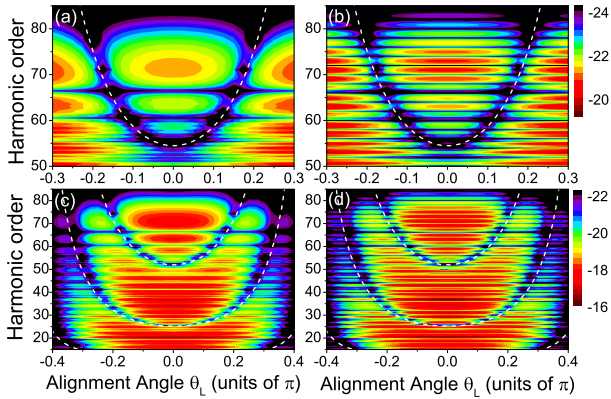}
\caption{(Color online) Spectra computed for $\mathrm{H}_2$ [panels (a) and (b)] and $\mathrm{Ar}_2$ [panels (c) and (d)] in a linearly polarized field ($\xi=0$), including the six shortest pairs of orbits starting in the first [panels (a) and (c)] and in both half cycles [panels (b) and (d)]. The field intensity and frequency have been chosen as $I=5 \times 10^{14}\mathrm{W}/\mathrm{cm}^2$ and $\omega=0.057$ a.u., respectively. The internuclear distances are $R^{(\mathrm{H}_2)}=1.4$ a.u and $R^{(\mathrm{Ar}_2)}=7.2$ a.u. The white dashed lines indicate the energy positions of the two-center interference minima. The yield is displayed in a logarithmic scale.}
\label{Fig7}
\end{figure}

\begin{figure*}[tbp]
\noindent\hspace*{-0.5cm}\includegraphics[scale=0.6]{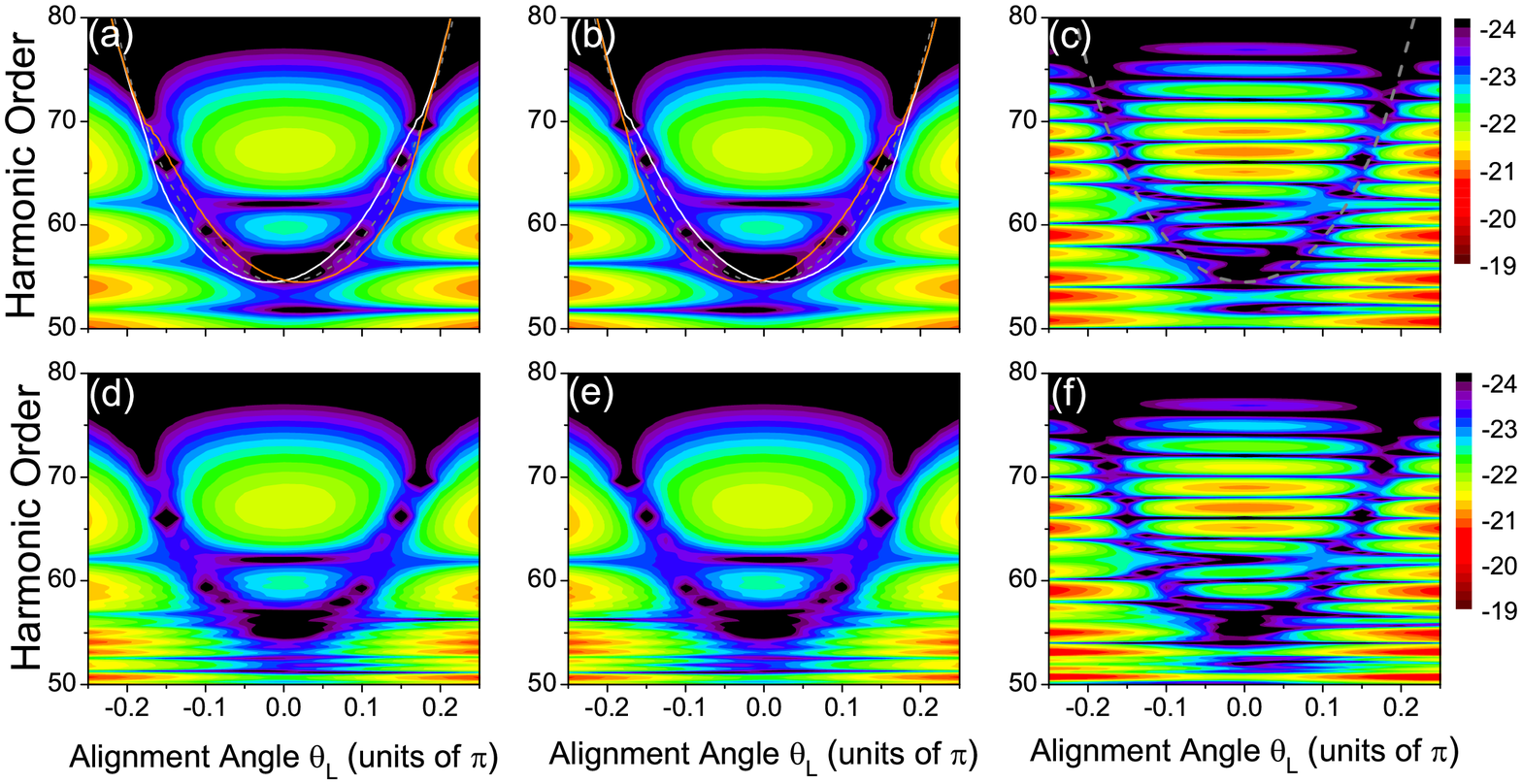}
\caption{Spectra computed for the same field and molecular parameters in
Fig. \protect\ref{Fig1} ($\phi=0.2$), but considering the coherent sums of the
transition amplitudes associated to: Orbits 1 and 2 [panel (a)]; orbits 1a and 2a [panel (b)]; orbits 1, 1a, 2 and 2a [panel (c)]; the three shortest pairs of orbits starting at the first half cycle; i.e., the pairs composed of orbits $(1,2)$, $(3,4)$ and $(5,6)$ [panel(d)]; the three shortest pairs of orbits 1a to 6a starting at the second half cycle [panel (e)]; the three shortest pairs of orbits, i.e., orbits 1 to 6a [panel (f)]. The interference conditions for the long and short orbits are given by the solid orange and white curves in panels (a) and (b), while the condition for linearly polarized fields is indicated by the dashed gray lines in panels (a), (b) and (c). The yield is displayed in a logarithmic scale.}
\label{Fig8}
\end{figure*}
In this section, we will study how the dynamic shifts discussed above will add up if a coherent
superposition of orbits is taken into account. This is important as, in a high-harmonic spectrum, there will be
several possibilities for the electron to return. Quantum mechanically, the
corresponding transition amplitudes will interfere, so that not only the real parts of such shifts,
but also their imaginary parts, become important. For comparison, we include the spectra computed for molecules in linearly polarized fields using the three shortest pairs of orbits. These spectra are displayed in Fig.~\ref{Fig7}, for $\mathrm{H}_2$ and $\mathrm{Ar}_2$ (upper and lower panels, respectively). The figure also shows other types of interference, that arise from the coherent superposition of ionization and recombination events displaced in time. In all panels, we notice that both the temporal interference patterns and the spatial, two-center interference minima are symmetric upon $\theta_L \rightarrow -\theta_L$. This is expected from our previous line of argument, and holds if orbits starting in the first half cycle [Figs.~\ref{Fig7}(a) and (c)], or in both half cycles [Figs.~\ref{Fig7}(b) and (d)] are included. Another noteworthy feature is the presence of well defined odd harmonics when the orbits starting at subsequent half cycles are added coherently, which can be clearly observed in Figs.~\ref{Fig7}(b) and (d). They are a consequence of the periodicity of the field, and are not present if the start times are restricted to the first half cycle.

In Fig.~\ref{Fig8}, we consider several coherent superpositions of orbits for elliptically polarized fields. We will first focus on the dominant pairs of orbits, i.e., $1$ and $2$, and, 1a and 2a, for $\mathrm{H}_2$ and $\phi=0.2$. These contributions are displayed in Figs.~\ref{Fig8}(a) and (b), together with the coherent superposition of the two dominant pairs [Fig.~\ref{Fig8}(c)]. These results are then compared to the spectra displayed in the lower panels of the figure, obtained using the three shortest pairs. Specifically, in Figs.~\ref{Fig8}(d), (e) and (f), we take orbits $1$ to $6$ starting in the first half cycle, orbits 1a to 6a starting in the second half cycle, and all six pairs of orbits, respectively.

All panels exhibit the u-shaped interference minimum, whose approximate position is roughly indicated by the interference conditions for linear and elliptical polarization (see the three curves in the figure). The outcomes of our simulations, however, do not follow a single interference curve. This is expected as the contributions from each orbit in a pair carry comparable weights, so that temporal interference effects between the long and short orbits play a role. The interference minima appear most clearly in the cutoff region, at roughly $\Omega=71\omega$, and at the bottom of the u-shaped minimum, near $\Omega=55\omega$. This is due to the fact that, in these energy regions, the interference conditions are closest. At the lower-energy end of the u-shaped minimum, the three interference curves cross. Hence, the two-center minimum is very visible. In the vicinity of this point, however, the three curves are very distinct. This implies that a blurring in the interference condition for a coherent superposition is expected in this region. At the cutoff, both $\mathrm{Re}[\zeta(t,t^{\prime})]$ and $\mathrm{Im}[\zeta(t,t^{\prime})]$ are closest and approach the interference condition for linear polarization. As a direct consequence, the two-center minimum is sharp around this frequency.  Beyond the cutoff, the imaginary parts of the shifts start to increase in absolute value and move away from each other. This will have little influence if only individual orbits are taken, as shown in the previous section, but will be critical for a coherent superposition of orbits. For that reason,
the minimum becomes blurred in this region. As in the linear case, there are high-order harmonics if orbits starting at different half cycles are included. The interference patterns, however, are no longer symmetric with regard to $\theta_L \rightarrow -\theta_L$, not even if the orbits starting in both cycles are taken into account [see Figs.~\ref{Fig8}(c) and (f)]. As expected, the spectra obtained for orbits starting at the second half cycle of the field, displayed in Figs.~\ref{Fig8}(b) and (e), are the mirror images of those computed using the orbits starting at the first half cycle, shown in Figs.~\ref{Fig8}(a) and (d). This holds not only for the u-shaped minimum, but also for the patterns associated with the interference of events displaced in time.

If the longer orbits are included, this leads at most to additional substructure in the low-plateau region, as a direct comparison of the lower and the upper panels of Fig.~\ref{Fig8} shows. This is caused by two main reasons. First, the excursion times of the electron in the continuum are much longer, in fact over one and a half cycles. Hence, a larger degree of wavepacket spreading occurs for the active electron, and this renders the contributions of such orbits less relevant. Second, the cutoff determined by such pairs is lower than that determined by the dominant orbits. In fact, for the parameters employed in the figure, it lies around $I_p+1.48U_p$ for orbits $3$ and $4$, and around $I_p+2.42U_p$ for orbits 5 and 6. This implies that, beyond harmonic frequencies $\Omega\simeq 55 \omega$, the contributions from such orbits are strongly suppressed. Finally, we observe an overall decrease in intensity, in comparison to the linearly polarized case. This is expected, as a nonvanishing ellipticity leads to a decrease in the tunnel ionization rate \cite{Delone_1991} and also in the return probability for the electron \cite{Dietrich_1994}.  There is also a displacement of the cutoff frequency towards lower energies, in agreement with previous studies in the literature \cite{Becker_1994,Wang_1999,Milosevic_2000}.

\begin{figure}[tbp]
\noindent\hspace*{-0.5cm}\includegraphics[scale=0.3]{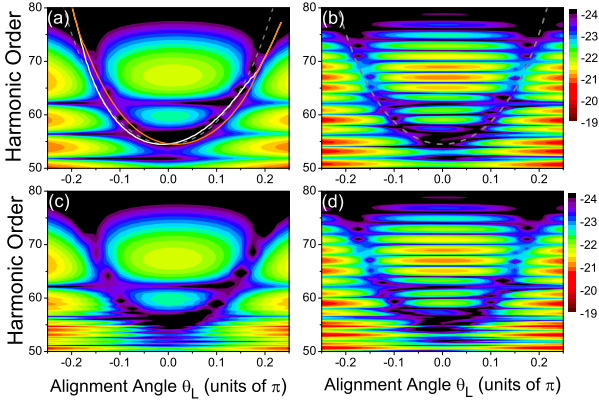}
\caption{(Color online) Spectra computed for the same field and molecular parameters in
Fig. \protect\ref{Fig3} ($\phi=0$), but considering the coherent sums of the
transition amplitudes associated to different combinations of orbits. Panels (a) and (b) include the dominant pair starting at the first half cycle and at both half cycles, respectively, while panels (c) and (d) include the contributions from orbits $1$ to $6$ and $1$ to 6a, respectively. The interference condition for linear polarization is indicated by the dashed lines in the upper panels, while its counterpart for elliptically polarized fields is given by the solid lines in panel 1. The orange and white lines refer to orbit 1 and 2, respectively. The yield is displayed in a logarithmic scale.}
\label{Fig9}
\end{figure}
In Fig.~\ref{Fig9}, we display the results obtained considering different coherent superpositions if both waves are in phase, i.e., for $\phi=0$. Also in this case, the main effect is a blurring of the structural interference condition, except at the lowest-energy part of the interference minimum and near the cutoff. An interesting aspect is how the residual shifts that exist beyond the cutoff behave.  If one considers start times in a specific subcycle, these shifts are apparent in the u-shaped structure. For instance, in Figs.~\ref{Fig9}(a) and (c), in which only orbits starting at the first half cycle have been included, one clearly sees that the suppression observed in the harmonic spectrum matches the solid lines in the cutoff region much more accurately than the interference condition for linear polarization (dashed gray line). Apart from that, this suppression is asymmetric and much more pronounced for $\theta_L>0$, i.e., on the right-hand side of these panels. This is in agreement with the previous discussions. If, however, the contributions from the first and second subcycles are added coherently, both this asymmetry and the residual shifts are washed out [see Figs~\ref{Fig9}(b) and (d)]. As expected from our previous discussion, (i) odd harmonics appear due to the periodicity of the field, as shown in the right panels, and (ii) the longer orbits do not influence the spectra considerably, as shown in the lower panels. An interesting effect is a blurring in the two-center minimum near the cutoff frequency (see harmonics $\Omega=65\omega$ to $\Omega=69\omega$) identified  in Figs.~\ref{Fig9}(b) and (d). This blurring is caused by the non-vanishing residual shifts from orbits located at different half cycles. These shifts are different for the orbits starting in the first and second half cycles, and smear the minimum if a coherent superposition is taken. For comparison, see Figs.~\ref{Fig8}(c) and (f), computed for $\phi=0.2$. As in this latter case the residual shift is vanishingly small near the cutoff, this blurring is absent.
\begin{figure}[tbp]
\noindent\hspace*{-0.5cm}\includegraphics[scale=0.3]{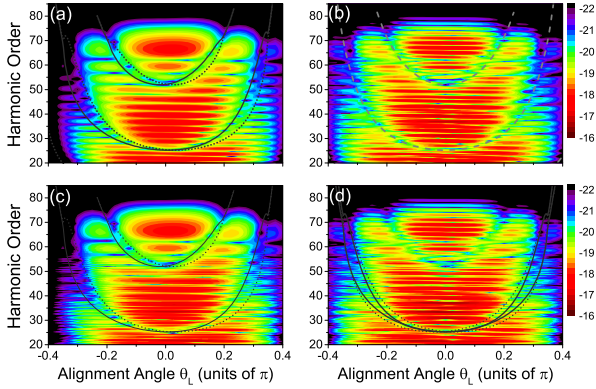}
\caption{(Color online) Spectra computed for argon using the elliptically polarized field of Fig.~\protect\ref{Fig1} ($\xi=0.3$, $\phi=0.2$) and different coherent superpositions of orbits. In panels (a) and (b), we included only the dominant orbits, while in panels (c) and (d) the six shortest pairs of orbits have been taken. In panels (a) and (c), we considered only ionization events starting in the first half cycle, while in panels (b) and (d) both first and second half cycles have been taken into consideration. The dotted and solid black lines in panels (a), (c) and (d) give the interference conditions for the long and short orbits, respectively. The dashed gray lines in panel (b) give the interference condition for linear polarization. In the figure, only the interference minima corresponding to $n=2$ and $n=3$ in Eq.~(\ref{intcon}) are visible. The yield is displayed in a logarithmic scale.}
\label{Fig10}
\end{figure}

In Fig.~\ref{Fig10} we exhibit the results computed for Ar$_2$ in an elliptically polarized field with $\phi=0.2$. We focus on the two-center minima $n=2$ and $n=3$ in Eq.~(\ref{intcon}). Apart from the above mentioned inaccuracies close to the threshold, inclusion of the minimum $n=1$ would require a much larger range of intensities and would obscure the effects we intend to analyze. The minimum $n=3$, located in the high-plateau region, behaves in a very similar way as that encountered for $\mathrm{H}_2$, i.e., there is an overall blurring with regard to the linearly polarized case and the minimum is clearest near the cutoff and at the bottom of the u-shaped minimum. The minimum $n=2$ spans a much larger harmonic region, so that the features observed are more dramatic. For this minimum, we no longer observe a structure as in Fig.~\ref{Fig7}, but a whole region in which suppression of the harmonic signal occurs, i.e., there is a splitting in the minimum. This region is bounded by the different interference conditions obtained for the long and short orbits, indicated by the solid and dotted lines in Figs.~\ref{Fig10}(a) and (c). This can be seen most clearly in Fig.~\ref{Fig10}(a), in which only the dominant orbits starting in the first half cycle has been included. This picture, however, persists if the longer pairs of orbits are included, as shown in Fig.~\ref{Fig10}(c). If the orbits starting in the second half cycle are also added coherently, this region will be bounded by the largest shifts $\mathrm{Re}[\zeta(t,t^{\prime})]$, which, in this case, correspond to the long orbits $1$ and 1a [dotted lines in Fig.~\ref{Fig10}(d)].
\begin{figure}[tbp]
\noindent\hspace*{-0.5cm}\includegraphics[scale=0.3]{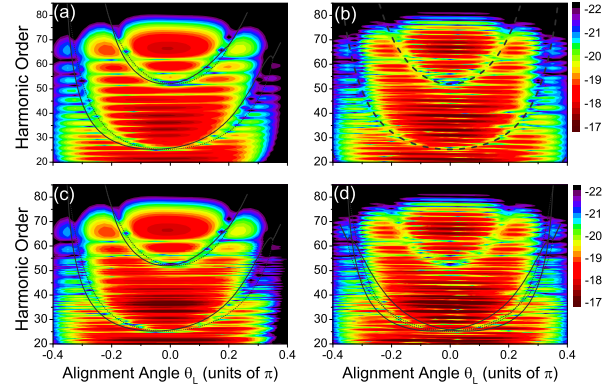}
\caption{(Color online) Spectra computed for argon using the same parameters and coherent superpositions of orbits as in Fig.~\protect\ref{Fig10}, but with a time delay of $\phi=0$. Panels (a) and (b): Dominant orbits (1 to 2 and 1 to 2a, respectively). Panels (c) and (d): six shortest pairs of orbits (1 to 6 and 1 to 6a, respectively). The dashed lines in panel (b) give the interference condition for linear polarization, and lines in the remaining panels give the interference condition (\ref{intcon}). The dotted and the solid lines refer to the long and short orbits, respectively. The yield is displayed in a logarithmic scale.}
\label{Fig11}
\end{figure}

Similar results, shown in Fig.~\ref{Fig11}, have been encountered for $\phi=0$. However, because of the residual shifts that exist for this phase, the splitting in the interference condition for the minimum $n=2$ is far more visible. This is specially true if the start times are restricted to a single half cycle, as shown in Fig.~\ref{Fig11}(a) and (c). In this latter case, there is also much larger asymmetry in the yield near the cutoff region for $n=2$. This is very visible if one compares the harmonic yield observed in the region $60<\Omega/\omega<70$ and alignment angle $\theta_L\simeq \pi/3$ with its counterpart for $\theta_L\simeq -\pi/3$. For the former angle, this yield is much more suppressed in this harmonic range. If however, one includes starting times in both half cycles [see Figs.~\ref{Fig11}(b) and (d)], this asymmetry is lost.
\section{Conclusions}
\label{conclusions}
In this paper, we have studied high-order harmonic generation in diatomic molecules in two-color elliptically polarized laser fields. We have shown that, even within a very simple model, namely the strong-field approximation and the single active electron, single active orbital approximation, a non-vanishing driving field ellipticity introduces a dynamic shift in a two-center interference condition which, for linear polarization, is purely structural. This shift depends very strongly on the orbit along which the active electron returns to its parent molecule, and on its kinetic energy upon return. What happens is that the angle with which the electron returns is effectively incorporated in the two center interference condition. Furthermore, depending on whether the electron returned with a non-vanishing transverse velocity at a field crossing, there may be a residual dynamic shift at and beyond the cutoff region for a given pair of orbits. A concrete example has been provided for the situation in which both low- and high-frequency driving waves were in phase.

For HHG transition probabilities related to individual orbits, we have found that, in general, our numerical results match very nicely the predictions from our generalized interference condition. If coherent superpositions of orbits are taken into account, the different shifts cause a blurring, and, in some cases, a splitting in the two-center minima. For elliptical polarization, these minima are no longer sharp, but, rather, there will be a region in the spectra for which the harmonic signal is suppressed. This region is bounded by the different interference conditions encountered for individual orbits. This splitting is also visible, though not discussed, in Ref.~\cite{Odzak_2010}.

Both the blurring and the splitting happen in most harmonic ranges, except in the cutoff region or when the modified interference conditions coincide. Hence, in a realistic situation, these dynamical shifts would mainly blur the two-center minima unless they converged to a single residual shift at the cutoff, or one of the orbits in a dominant pair could be suppressed. For instance, a clear asymmetry has been found for the minimum $n=2$ in Fig.~\ref{Fig11} if only the events starting in the first half cycle are considered.  In practice, this could be realized by suppressing the events in subsequent half cycles by an adequate field choice, such as, for instance, with a few-cycle pulse.
\section*{Acknowledgements}
We would like to thank X. Lai for very useful discussions. This work has been funded by the UK EPSRC (grant EP/J019240/1 and doctoral training prize) and by UCL (Impact Studentship).

%\bibliographystyle{prsty}
%\bibliography{ReviewCarla2}

\end{document}